# Do evidence-based active-engagement courses reduce the gender gap in introductory physics?

Nafis I. Karim[1], Alexandru Maries[2] and Chandralekha Singh[1]

[1] *Department of Physics and Astronomy, University of Pittsburgh, Pittsburgh, PA 15260*
[2] *Department of Physics, University of Cincinnati, Cincinnati, OH 45221*

**Abstract.** Prior research suggests that using evidence-based pedagogies can not only improve learning for all students, it can also reduce the gender gap. We describe the impact of physics education research based pedagogical techniques in flipped and active-engagement non-flipped courses on the gender gap observed with validated conceptual surveys. We compare male and female students' performance in courses which make significant use of evidence-based active engagement (EBAE) strategies with courses that primarily use lecture-based (LB) instruction. All courses had large enrollment and often had more than 100 students. The analysis of data for validated conceptual surveys presented here includes data from two-semester sequences of introductory algebra-based and calculus-based introductory physics courses. The conceptual surveys used to assess student learning in the first and second semester courses were the Force Concept Inventory and the Conceptual Survey of Electricity and Magnetism, respectively. In the research discussed here, the performance of male and female students in EBAE courses at a particular level is compared with LB courses in two situations: (I) the same instructor taught two courses, one of which was an EBAE course and the other an LB course, while the homework, recitations and final exams were kept the same, (II) student performance in all of the EBAE courses taught by different instructors was averaged and compared with LB courses of the same type also averaged over different instructors. In all cases, on conceptual surveys we find that students in courses which make significant use of active-engagement strategies, on average, outperformed students in courses of the same type using primarily lecture-based instruction even though there was no statistically significant difference on the pretest before instruction. However, the gender gap persisted even in courses using EBAE methods. We also discuss correlations between the performance of male and female students on the validated conceptual surveys and the final exam, which had a heavy weight on quantitative problem solving.

## I. INTRODUCTION

### A. Physics Education Research-based Active Engagement Methods

In the past few decades, physics education research has identified challenges that students encounter in learning physics at all levels of instruction [1-7]. Building on these investigations, researchers are developing, implementing and evaluating evidence-based curricula and pedagogies to reduce these challenges to help students develop a coherent understanding of physics concepts and enhance their problem solving, reasoning and metacognitive skills [8-18]. In evidence-based curricula and pedagogies, the learning goals and objectives, instructional design, and assessment of learning are aligned with each other and there is focus on evaluating whether the pedagogical approaches employed have been successful in meeting the goals and enhancing student learning.

One highly successful model of learning is the field-tested cognitive apprenticeship model [19]. According to this model, students can learn effectively if the instructional design involves three essential components: "modeling", "coaching and scaffolding", and "weaning". In this approach, "modeling" means that the instructional approaches demonstrate and exemplify the criteria for good performance and the skills that students should learn (e.g., how to solve physics problems systematically). "Coaching and scaffolding" means that students receive guidance and support as they actively engage in learning the content and skills necessary for good performance. "Weaning" means gradually reducing the support and feedback to help students develop self-reliance [19]. In traditional physics instruction, especially at the college level, there is often a lack of coaching and scaffolding: students come to class where the instructor lectures and does some example problems, then students are left on their own to work through homework with little or no feedback. This lack of prompt feedback and scaffolding can be detrimental to learning.

Some of the commonly used evidence-based active-engagement (EBAE) approaches implemented in physics include peer instruction with clickers popularized by Eric Mazur from Harvard University [20-22], tutorial-based instruction in introductory and advanced courses [23-25] and collaborative group problem solving [26-29], e.g., using context-rich problems [4-5]. In all of these evidence-based approaches, formative assessment plays a critical role in student learning [30]. Formative assessment tasks are frequent, low-stakes assessment activities which give feedback to students and instructors about what students have learned at a given point. Using frequent formative assessments helps make the learning goals of the course concrete to students, and provides them with a way to track their progress in the course with respect to these learning goals. When formative assessment tasks such as concept-tests, tutorials and collaborative group problem solving are interspersed throughout the course, learning is enhanced [30-32].

Moreover, technology is increasingly being exploited for pedagogical purposes to improve student learning. For example, Just-in-Time Teaching (JiTT) is an instructional approach in which instructors receive feedback from students before class and use that feedback to tailor in-class instruction [33-34]. Typically, students complete an electronic pre-lecture assignment in which they give feedback to the instructor regarding any difficulties they have had with the assigned reading material, lecture videos, and/or other self-paced instructional tools. The instructor then reviews student feedback before class and makes adjustments to the in-class activities. For example, during class, the instructor can focus on student difficulties found via electronic feedback. Students may engage in discussions with the instructor and with their classmates, and the instructor may then adjust the next pre-lecture assignment based on the progress made during class. When JiTT was first conceived and implemented in the late 1990s in physics classes, the required internet technology for electronic feedback was still evolving; developments in digital technology since then have continued to make electronic feedback from students and the JiTT approach easier to implement in classes. For example, Eric Mazur's Perusall system [35] allows students to read the textbook and ask questions electronically and the system uses their questions to draft a "confusion report" which distills their questions to three most common difficulties, which can be addressed in class. It has been hypothesized that JiTT may help students learn better because out-of-class activities cause students to engage with and reflect on the parts of the instructional material they find challenging [33-34]. In particular, when the instructor focuses on student difficulties in lecture which were found via electronic feedback before class, it may create a "time for telling" [36] especially because students may be "primed to learn" better when they come to class if they have struggled with the material during pre-lecture activities. The JiTT approach is often used in combination with peer discussion and/or collaborative group problem solving inter-dispersed with lectures in the classroom.

In addition, in the last decade, the JiTT pedagogy has been extended a step further with the maturing of technology [37-42] and "flipped" [43,44] classes with limited in-class lectures have become common with instructors asking students to engage with short lecture videos (or read certain section of the textbook) and concept questions associated with each video outside of the class and using most of the class-time for active-engagement. The effectiveness of flipped classes in enhancing student learning can depend on many factors including the degree to which evidence-based pedagogies that build on students' prior knowledge and actively engage them in the learning process are used, whether there is sufficient buy-in from students, and the incentives that are used to get students engaged with the learning tools both inside and outside the classroom.

Moreover, research suggests that effective use of peer collaboration can enhance student learning in many instructional settings in physics classes, including in JiTT and flipped environments, and with various types and levels of student populations. Although the details of implementation vary, students can learn from each other in many different environments. For example, in Mazur's peer instruction approach [45], the instructor poses concrete conceptual problems in the form of conceptual multiple-choice clicker questions to students throughout the lecture and students discuss their responses with their peers. Heller et al. have shown that collaborative problem solving with peers in the context of quantitative "context-rich" problems [4-5] can be valuable both for learning physics and for developing effective problem solving strategies.

In evidence-based "active-engagement non-flipped" courses [46], lecture and interactive activities are combined during the prescribed class time to enhance student learning and students' out-of-class homework assignments are often similar to those assigned in traditionally taught classes. On the other hand, in flipped courses, there is very limited direct instruction (lecture) and the majority of in-class time is used to actively engage students in learning. The effectiveness of flipped classes depends on how the course is designed and incentivized and how out-of-class activities build on in-class activities. In addition, whether instructors create a low or high anxiety active-learning environment can play a critical role in student engagement. It can particularly impact learning for women and students from other underrepresented groups, whose sense of belonging and self-efficacy can either be enhanced or exacerbated depending upon the design of the active-learning environment. More about these types of issues is discussed in the next section.

Also, the lecture videos that students often watch outside of the class in a flipped class are self-paced, which has both advantages and disadvantages. While pedagogically developed, implemented and incentivized self-paced videos can provide a variety of students with an opportunity to learn at a pace that is commensurate with their prior knowledge, without appropriate pedagogy in the development, implementation, and incentives to learn from these tools, students may not engage with them as intended, especially if they do not have good time-management and self-regulation skills. For example, research on Massive Open Online Courses (MOOCs) [47] suggests that a majority of those who complete the entire online course already have a bachelor's degree. Moreover, a student who does not keep up with out-of-class activities such as watching videos and answering the concept questions associated with them before coming to class is unlikely to take full advantage of the interactive in-class activities in a flipped class. Thus, while a well-designed and implemented flipped course has the potential to help a variety of students learn to think like a physicist and can scaffold their learning of physics, many students may not engage and learn from the out-of-class videos if they are not intrinsically motivated and if the videos are not effective [37] or are not implemented and incentivized appropriately. Despite these caveats, well-designed and well-implemented interactive videos [48] and associated questions designed carefully can be beneficial as they can help a variety of students with different prior preparations and allow them to learn at their own pace. Moreover, if the videos are part of an adaptive video-suite for students with different prior knowledge and skills (for example, after a student views a video, he/she can be asked several questions, and if he/she struggles to answer those questions, he/she can be directed to another explanation video that other students who answer those questions correctly can skip). In particular, the videos can provide more scaffolding support as needed to a student who is struggling. Then, after taking full advantage of these out-of-class activities, the EBAE activities can help all students.

## B. Gender gap in introductory physics courses

Prior research has found that male students outperform female students on standardized conceptual assessments such as the Force Concept Inventory (FCI) [49] or the Conceptual Survey of Electricity and Magnetism (CSEM) [50]. The discrepancy between male and female students' performance is typically referred to as a "gender gap" [51-53]. While sometimes gender gap can be accounted for at least in part due to different prior preparation or coursework of male and female students, it has also been found even after controlling for these factors [50]. Prior research has also found that using evidence-based pedagogies can reduce the gender gap [54-55], but the extent to which this occurs varies. Others have found that the gender gap is not reduced despite significant use of evidence-based pedagogies [56]. Prior research has also found a gender gap on other assessments such as a conceptual assessment for introductory laboratories [57] and physics exams [51-52]. Yet others have found no differences in performance between male and female students on exams [53,58-59].

The origins of gender gap on the FCI both at the beginning and end of a physics course have been a subject of debate with some researchers arguing that the test itself is gender-biased [60]. Some of the origins of the gender gap are related to societal gender stereotypes [61-64] that keep accumulating from an early age. For example, research suggests that even six year old boys and girls have gendered views about smartness in favor of boys [64]. Such stereotypes can impact female students' self-efficacy [65,66], their beliefs about their ability to perform well, in disciplines such as physics in which they are underrepresented and which have been associated with "brilliance". They can also impact their intelligence mindset [67], which is related to beliefs about whether intelligence innate or whether it is something that can be developed and cultivated via focus and persistence in problem solving in a discipline such as physics. Thus, it may not be surprising that prior research has found that activation of a stereotype, i.e., stereotype threat (ST) about a particular group in a test-taking situation can alter the performance of that group in a way consistent with the stereotype [61-64]. In fact, some researchers have argued [61] that female students, when working on a physics assessment, undergo an implicit ST due to the prevalent societal stereotypes. In particular, Marchand and Taasoobshirazi [61] conducted a study in which high school students were randomly divided into three groups and all students received the following instructions before taking a physics test: "You will be given four physics problems to solve. These problems are based on physics material that you have already covered." In the implicit ST condition, these were the only instructions, while in the explicit ST condition, students were also told: "This test has shown gender differences with males outperforming females on the problems" and in the nullified condition, students were told: "No gender differences in performance have been found on the test." They found no statistically significant difference on the physics test between female students' performance in the explicit ST condition and the implicit ST condition but female students in both these conditions performed significantly worse than male students. In contrast, the nullified condition in which female students were instead told that the test they are about to take is gender neutral erased the gender gap (no difference in performance between male and female students). The researchers hypothesized [61] that simply administering a physics test to female students creates an implicit stereotype threat (which is partly due to societal gender bias and related issue of anxiety and self-efficacy, which refers to the fact that many female students start doubting their own ability to perform well in a physics test).

**C. Focus of our research**

In this study, we used the FCI [49] in the first semester introductory physics courses and the CSEM [50] in the second semester courses to assess student learning. We also investigated any possible gender gap at the beginning of the course as well as the extent to which evidence-based pedagogies can help reduce it. The FCI, CSEM and other standardized physics surveys [68-73] have been used to assess introductory students' understanding of physics concepts by a variety of educators and physics education researchers. One reason for their extensive use is that many of the items on the survey have strong distractor choices which correspond to students' common difficulties so students are unlikely to answer the survey questions correctly without having good conceptual understanding. Our research focuses on the following research questions for both algebra-based and calculus-based introductory physics courses:

**RQ1.** What is the gender gap on the FCI/CSEM pretest and posttest in LB and EBAE courses? By how much do both male and female students improve from pretest to posttest in LB and EBAE courses?

**RQ2.** How does the performance on the FCI/CSEM of male and female students in LB courses compare to EBAE courses in both the pretest and the posttest?

**RQ3.** To what extent do male and female students with high or low pretest scores perform differently in EBAE courses compared to LB courses when the comparison is made for one instructor who teaches both an EBAE and an LB course at the same time?

**RQ4.** To what extent do male and female students with high or low pretest scores perform differently in EBAE courses compared to LB courses when the comparison is made between EBAE and LB courses taught by different instructors?

**RQ5.** Is there any correlation between posttest and final exam scores for male and female students?

Thus, in our research, the performances of male and female students in EBAE courses in a particular type of course (algebra-based or calculus-based physics I or II) are compared with male and female students of LB courses in two situations: (I) the same instructor taught two courses, one of which was an EBAE course and the other an LB course with common homework and final exams, (II) student performances in all of the EBAE courses taught by different instructors were averaged and compared with LB courses of the same type, also averaged over different instructors.

Also, the students were divided into three subgroups based upon their pretest scores: top 1/3rd, middle 1/3rd and bottom 1/3rd. We calculated whether there was a statistically significant difference between male and female students' average scores on the pretest, posttest or final exam in two cases: (i) male students were divided into three subgroups according to the pretest scores of males only and female students were also divided into three subgroups according to the pretest scores of females only, and then the male and female students' average scores in each subgroup were compared and (ii) all students were divided into the three subgroups according to their pretest scores *regardless* of their gender and then male and female students in each of the three subgroups were separated and compared. This type of analysis based upon gender was carried out for the male and female students taught by the same instructor (teaching either LB or EBAE course) and also for different instructors teaching LB or EBAE courses of the same type combined. Whenever differences between these two groups were observed (e.g., with male or female students in the EBAE courses on average performing better than the corresponding students in the LB courses), we investigated which subgroup was benefiting most from the EBAE courses, e.g., those who performed well or poorly on the pretest given at the beginning of the course. Finally, we investigated the typical correlation between the performance of male and female students' posttest performances on the validated conceptual surveys and their performance on the instructor-developed final exam (which typically places a heavy weight on quantitative physics problems).

## II. METHODOLOGY

### A. Courses and Participants

The participants in this study were students in 16 different algebra-based and calculus-based introductory physics courses. Out of all introductory physics courses (algebra-based or calculus-based physics I or II) included in this study, there were four EBAE courses: two completely flipped classes in algebra-based introductory physics I and one completely flipped and one interactive active engagement class in calculus-based introductory physics II. These courses include approximately 700 male and 750 female students in first semester courses and approximately 650 male and 500 female students in second semester courses at a typical large research university in the US (University of Pittsburgh). The details of the courses that fall into three categories are as follows:

1) A lecture-based (or LB) course is one in which the primary mode of instruction was via lecture. In addition to the three or four weekly hours for lectures, students attended an hour long recitation section taught by a graduate TA. In recitation, the TA typically answered student questions (mainly about their homework problems which were mostly textbook style quantitative problems), solved problems on the board and gave students a quiz in the last 10-20 minutes.

2) A flipped course is one in which the class was broken up into two almost equal size groups with each group meeting with the instructor for half the regular class time. For example, for a 200 student class scheduled to meet for four hours each week (on two different days), the instructor met with half the class (100 students) on the first day and the other half on the second day. This was possible in the flipped classes since the total contact hours for each instructor each week with the students was the same as in the corresponding LB courses. Students watched the lecture videos before coming to class and answered some conceptual questions which were based upon the lecture video content. They uploaded the answers to those conceptual questions before class onto the course website and were scored for a small percentage of their grade (typically 4-8%). Although students had to watch several videos outside of class in preparation for each class, each video was typically 5-10 minutes long, followed by concept questions. On average, students in a flipped class had to watch recorded videos which took a little less than half the allotted weekly time for class (e.g., for the courses scheduled for four hours each week, students watched on average 1.5 hours of videos each week, and in the courses scheduled for three hours each week, students watched around one hour of videos). These video times do not include the time that students would take to rewind the video, stop and think about the concepts and answer the concept questions placed after the videos that counted towards their course grade. In the spirit of JiTT, the instructors of the flipped courses adjusted the in-class activities based upon student responses to online concept questions which were supposed to be submitted the night before the class. About 90% of the students submitted their answers to the concept questions that followed the videos to the course website before coming to the class. The web-platforms used for managing, hosting and sharing these videos and for having online discussions with students about them asynchronously (in which students and the instructor participated) were Classroom Salon or Panopto. In-class time was used for clicker questions involving peer discussion and then a whole class discussion of the clicker questions, collaborative group problem solving involving quantitative problems in which 2-3 students worked in a group (followed by a clicker question about the order of magnitude of the answer), and lecture-demonstrations with preceding clicker questions on the same concepts. In addition to the regular class times, students attended an hour long recitation section which was taught the same way as for students in the LB courses.

It is important to note that the instructors who taught the flipped courses also taught LB courses at the same time (usually teaching two courses in a particular semester: one flipped and one LB). Students in both flipped and LB courses completed the same homework and took the same final exam. For the calculus-based flipped courses, the students also took the same midterm exams. This was not possible for the algebra-based courses because the exams were scheduled at different times. However, in the algebra-based courses they took the same final exam and had the same homework.

3) In an EBAE interactive non-flipped course, the instructor combined lectures with research-based pedagogies including clicker questions involving peer discussion, conceptual tutorials, collaborative group problem solving, and lecture demonstrations with preceding clicker questions on the same concepts, similar to the flipped courses. In addition, students attended a reformed recitation which primarily used context-rich problems to get students to engage in group problem solving or worked on research-based tutorials while being guided by a TA. The instructor ensured that the problems students solved each week in the recitation activities were closely related to what happened in class. Students also worked on some research-based tutorials during class in small groups, but if they did not finish them in the allotted time, they were asked to complete them at home and submit as homework.

From now on, we refer to the flipped and interactive non-flipped courses as EBAE courses except when relevant. We also note that the number of female students in algebra-based courses is larger than that of male students. Most of the algebra-based students have biological science or related majors like Biology, Psychology, Exercise Science, Neurology/Neuromedicine, Environmental Science, etc. In calculus-based courses, on the other hand, there are more male than female students. Most calculus-based students are in their first year in college, and have physical science related majors such as chemistry, mathematics, engineering (electrical, mechanical, chemical, civil etc.), and physics (typically only 5-10 physics majors out of several hundred students). The algebra-based or calculus-based physics courses are mandatory for these students. We do not have information about the background of the students, such as their prior experiences in physics or mathematics before college or whether they took any physics or math courses in high school (although a majority of these students have typically taken at least one high school physics course and the typical percentage of female students in calculus-based "advanced placement C" high school courses in the US is less than one third).

We also note that none of the instructors teaching the EBAE courses focused explicitly on whether the active-learning classroom environment helped foster a sense of belonging or focused on improving self-efficacy and instilling a growth mindset in all students. In particular, the instructors did not *explicitly* focus on whether the active-learning classroom was a low anxiety classroom for all students and whether women and other underrepresented students felt supported and had the same level of engagement with the active-learning activities.

## B. Materials

The materials used in this study are the FCI and CSEM conceptual multiple-choice (five choices for each question) standardized surveys, which were administered in the first week of classes before instruction in relevant concepts (pretest) and after instruction in relevant concepts (posttest). The FCI was used in the first semester courses and the CSEM was used in the second semester courses. Apart from the data on these surveys that the researchers collected from all of these courses, each instructor administered his/her own final exam, which was mostly quantitative (60%-90% of the questions were quantitative, although some instructors had either the entire final exam or part of it in a multiple-choice format with five options for each question to make grading easier). Ten course instructors (who also provided the FCI or CSEM data from their classes) provided their students' final exam scores and most of them also provided a copy of their final exams.

## C. Methods

Our main goals in this research were to compare the average performances of male and female students in introductory physics courses in different types of classes (e.g., Algebra-based or Calculus-based, EBAE or LB) and to compare male and female students' performances between courses that used EBAE pedagogies with the performances of students in LB courses by using standardized conceptual surveys, the FCI (for physics I) and CSEM (for physics II) as pre/posttests. We not only calculated the average gain (posttest - pretest scores) for each group for males and females but also calculated the average normalized gain, which is commonly used to determine how much students learned from pretest to posttest taking into account their initial scores on the pretest, to find out whether the gender gap increased, decreased or remained the same. The normalized gain is defined as $\langle g \rangle = \frac{\%\langle S_f \rangle - \%\langle S_i \rangle}{100 - \%\langle S_i \rangle}$, in which $\langle S_f \rangle$ and $\langle S_i \rangle$ are the final (post) and initial (pre) class averages, respectively. Then, $Norm\ g = 100\langle g \rangle$ in percent [16]. This normalized gain provides valuable information about how much students have learned by taking into account what they already know based on the pretest. We wanted to investigate whether the normalized gain is higher in one course compared to another, and whether it is the same or different for males and females.

In order to compare EBAE courses with LB courses, we performed $t$-tests [74] on FCI or CSEM pre and posttest data for males and females. We also calculated the effect size in the form of Cohen's $d$ defined as $(\mu_1 - \mu_2)/\sigma_{pooled}$, where $\mu_1$ and $\mu_2$ are the averages of the two groups being compared (e.g., EBAE vs. LB or male vs. female) and $\sigma_{pooled} = \sqrt{(\sigma_1{}^2 + \sigma_2{}^2)/2}$ (here $\sigma_1$ and $\sigma_2$ are the standard deviations of the two groups being compared). We considered: $d < 0.5$ as small effect size, $0.5 \leq d < 0.8$ as medium effect size and $d \geq 0.8$ as large effect size, as described in Ref. [75].

Moreover, although we did not have control over the type of final exam each instructor used in his/her courses, we wanted to look for correlations between the FCI/CSEM posttest performance and the final exam performance for different instructors in the algebra-based and calculus-based EBAE or LB courses for male and female students separately. Including both the algebra-based and calculus-based courses, 10 instructors provided the final exam scores for their classes. We used these data to obtain linear regression plots between the posttest and the final exam performance for males, females and all students (combined) for each instructor and computed the correlation coefficient between the performance of students (male/female/all) on the validated conceptual surveys and their performance on the final exam for different instructors. These correlation coefficients between the conceptual surveys and the final exam (with strong focus on quantitative problem solving) can provide an indication of the strength of the correlation between conceptual and quantitative problem solving of male and female students in these courses.

## III. RESULTS

**RQ1. Comparison of the gender gap on the FCI/CSEM pretest and posttest in LB and EBAE courses**

    **Physics I:** In Table I, we present intra-group pre/posttest data (pooled data for the same type of courses) of male and female students on the FCI for the calculus-based and algebra-based introductory physics I courses. For the algebra-based courses, some were EBAE courses while others were LB courses, whereas all the calculus-based courses were LB. We found statistically significant improvements from the pretest to the posttest for each group (for both male and female students) in both LB and EBAE courses. However,

Table I. Intra-group FCI pre/posttest averages (Mean) and standard deviations (SD) for first-semester introductory male and female students in calculus-based LB courses, and algebra-based EBAE and LB courses. The number of students in each group, N, is shown. For each group, a *p*-value obtained using a *t*-test shows that the difference between the pre/posttest is statistically significant and the difference between the male and female students is also statistically significant. The normalized gain (Norm g) from pretest to posttest and the effect size (Eff. size) shows how much male and female students learned from what they did not already know based on the pretest.

| Type of Class | FCI | Female | Gender Comparison | Male |
|---|---|---|---|---|
| Calc LB | Pretest | N: 146<br>Mean: 43%<br>SD: 20% | p-value: <0.001<br>← **gender gap: 13%** →<br>Eff. size: 0.635 | N: 283<br>Mean: 55%<br>SD: 20% |
| | Pretest vs. Posttest Comparison | ↑<br>p-value: <0.001<br>Eff. size: 0.468<br>Norm g: 16%<br>↓ | | ↑<br>p-value: <0.001<br>Eff. size: 0.686<br>Norm g: 30%<br>↓ |
| | Posttest | N: 114<br>Mean: 52%<br>SD: 20% | p-value: <0.001<br>← **gender gap: 17%** →<br>Eff. size: 0.868 | N: 200<br>Mean: 68%<br>SD: 19% |
| Alg EBAE | Pretest | N: 153<br>Mean: 30%<br>SD: 15% | p-value: <0.001<br>← **gender gap: 13%** →<br>Eff. size: 0.749 | N: 105<br>Mean: 43%<br>SD: 20% |
| | Pretest vs. Posttest Comparison | ↑<br>p-value: <0.001<br>Eff. size: 1.176<br>Norm g: 28%<br>↓ | | ↑<br>p-value: <0.001<br>Eff. size: 0.955<br>Norm g: 32%<br>↓ |
| | Posttest | N: 149<br>Mean: 49%<br>SD: 18% | p-value: <0.001<br>← **gender gap: 12%** →<br>Eff. size: 0.653 | N: 106<br>Mean: 61%<br>SD: 19% |
| Alg LB | Pretest | N: 456<br>Mean: 30%<br>SD: 14% | p-value: <0.001<br>← **gender gap: 11%** →<br>Eff. size: 0.691 | N: 318<br>Mean: 41%<br>SD: 18% |
| | Pretest vs. Posttest Comparison | ↑<br>p-value: <0.001<br>Eff. size: 0.930<br>Norm g: 21%<br>↓ | | ↑<br>p-value: <0.001<br>Eff. size: 0.863<br>Norm g: 28%<br>↓ |
| | Posttest | N: 383<br>Mean: 44%<br>SD: 18% | p-value: <0.001<br>← **gender gap: 13%** →<br>Eff. size: 0.686 | N: 255<br>Mean: 57%<br>SD: 20% |

both female and male students exhibited larger normalized gains in the EBAE courses. In the calculus-based LB course, the gender gap increased slightly from 13% to 17%, whereas in the algebra-based courses, the gender gap stayed roughly the same (varied between 11% and 13%) both in LB and EBAE courses. In both the pretest and the posttest in calculus-based and algebra-based courses, the difference in performance between male and female students was statistically significant and the effect sizes were typically in the medium range. Thus, it appears that in algebra-based courses, using evidence-based pedagogies helped both female and male students learn more, but did not result in a reduction of the gender gap.

Table II. Intra-group CSEM pre/posttest averages (Mean) and standard deviations (SD) for second-semester introductory male and female students in calculus-based LB and EBAE courses and algebra-based LB courses. The total number of students in each group, N, is shown. For each group, a *p*-value obtained using a *t*-test shows that the difference between the pre/posttest is statistically significant and the difference between the male and female students is also statistically significant. The normalized gain (Norm g) from pretest to posttest and the effect size (eff. Size) shows how much male and female students learned from what they did not already know based on the pretest.

| Type of Class | CSEM | Female | Gender Comparison | Male |
|---|---|---|---|---|
| Calc LB | Pretest | N: 84<br>Mean: 34%<br>SD: 13% | p-value: 0.007<br>← **gender gap: 4%** →<br>Eff. size: 0.349 | N: 234<br>Mean: 38%<br>SD: 13% |
| | Pretest vs. Posttest Comparison | ↑<br>p-value: <0.001<br>Eff. size: 0.821<br>Norm g: 18%<br>↓ | | ↑<br>p-value: <0.001<br>Eff. size: 0.894<br>Norm g: 22%<br>↓ |
| | Posttest | N: 78<br>Mean: 45%<br>SD: 16% | p-value: 0.003<br>← **gender gap: 6%** →<br>Eff. size: 0.381 | N: 248<br>Mean: 51%<br>SD: 17% |
| Calc EBAE | Pretest | N: 112<br>Mean: 35%<br>SD: 14% | p-value: 0.017<br>← **gender gap: 4%** →<br>Eff. size: 0.272 | N: 220<br>Mean: 39%<br>SD: 16% |
| | Pretest vs. Posttest Comparison | ↑<br>p-value: <0.001<br>Eff. size: 1.143<br>Norm g: 28%<br>↓ | | ↑<br>p-value: <0.001<br>Eff. size: 1.384<br>Norm g: 39%<br>↓ |
| | Posttest | N: 98<br>Mean: 53%<br>SD: 18% | p-value: <0.001<br>← **gender gap: 10%** →<br>Eff. size: 0.538 | N: 193<br>Mean: 63%<br>SD: 19% |
| Alg LB | Pretest | N: 301<br>Mean: 22%<br>SD: 8% | p-value: <0.001<br>← **gender gap: 6%** →<br>Eff. size: 0.452 | N: 201<br>Mean: 27%<br>SD: 13% |
| | Pretest vs. Posttest Comparison | ↑<br>p-value: <0.001<br>Eff. size: 1.450<br>Norm g: 23%<br>↓ | | ↑<br>p-value: <0.001<br>Eff. size: 1.325<br>Norm g: 29%<br>↓ |
| | Posttest | N: 266<br>Mean: 40%<br>SD: 16% | p-value: <0.001<br>← **gender gap: 8%** →<br>Eff. size: 0.451 | N: 172<br>Mean: 48%<br>SD: 18% |

**Physics II:** In Table II, we present intra-group (pooled data for the same type of courses) pre/posttest data for male and female students on the CSEM survey for algebra-based and calculus-based introductory physics II courses. Similar to the data shown in Table I, we found statistically significant improvements on the CSEM for female and male students both in LB and EBAE courses, however, the learning gains for both female and male students were larger in EBAE courses. With regards to the gender gap, we found that in LB courses it stayed roughly the same (4% on pretest and 6% on posttest for calculus-based courses, 6% on the pretest and 8% on posttest for algebra-based courses). However, in the EBAE calculus-based course, the gender gap increased slightly from 4% to 10%, and it appears that male students may have benefited more from evidence-based pedagogies than female students (normalized gain for male students was 39% in EBAE courses compared to 29% for female students).

**RQ2. Comparison of the performance on the FCI/CSEM of male and female students in LB and EBAE courses in both the pretest and the posttest**

**Physics I:** Table III shows the between-course male and female student FCI pre/posttest score comparison between algebra-based LB and EBAE courses, first holding the instructor fixed (same instructor taught both the LB and EBAE courses, used the same homework and final exams) and second, combining all instructors who used similar methods in the same group (only one instructor used EBAE methods, but several who taught LB courses were combined). Table III shows that on the pretest, the performance of male and female students in the LB courses was similar to the EBAE courses. However, on the posttest both male (female) students in the EBAE courses outperformed male (female) students in the LB courses (effect sizes ranging from 0.196 to 0.324). Coupled with the gender gaps shown in Table I, these data suggest that while both female and male students learned more in EBAE courses, the gender gap remained roughly the same in EBAE courses as in LB courses.

Table III. Between-course comparison of the average FCI pre/posttest scores of algebra-based male and female students in LB courses with EBAE courses when (i) both courses are taught by the same instructor and (ii) different instructors using similar instructional methods are combined. The p-values and effect sizes are obtained for male and female students separately when comparing the LB and EBAE courses in terms of students' FCI scores.

| | FCI | Female-Pretest | Female-Posttest | Male-Pretest | Male-Posttest |
|---|---|---|---|---|---|
| (i) FCI Alg: LB vs. EBAE (same instructor) | LB | N: 260 Mean: 30% SD: 13% | N: 246 Mean: 43% SD: 18% | N: 166 Mean: 42% SD: 19% | N: 154 Mean: 56% SD: 21% |
| | LB vs. EBAE Comparison | p-value: 0.846 Eff. size: 0.020 | p-value: 0.002 Eff. size: 0.324 | p-value: 0.786 Eff. size: 0.034 | p-value: 0.041 Eff. size: 0.258 |
| | EBAE | N: 153 Mean: 30% SD: 15% | N: 149 Mean: 49% SD: 18% | N: 105 Mean: 43% SD: 20% | N: 106 Mean: 61% SD: 19% |
| (ii) FCI Alg: LB vs. EBAE (different instructors combined) | LB | N: 456 Mean: 30% SD: 14% | N: 383 Mean: 44% SD: 18% | N: 318 Mean: 41% SD: 18% | N: 255 Mean: 57% SD: 20% |
| | LB vs. EBAE Comparison | p-value: 0.861 Eff. size: 0.017 | p-value: 0.009 Eff. size: 0.255 | p-value: 0.432 Eff. size: 0.091 | p-value: 0.088 Eff. size: 0.196 |
| | EBAE | N: 153 Mean: 30% SD: 15% | N: 149 Mean: 49% SD: 18% | N: 105 Mean: 43% SD: 20% | N: 106 Mean: 61% SD: 19% |

**Physics II:** Table IV shows the between-course CSEM pre/posttest score comparison between calculus-based LB and EBAE courses, first holding the instructor fixed (same instructor taught both the LB and EBAE courses and used the same homework and final exams) and second, combining all instructors who taught using similar methods into the same group. Table IV shows that on the pretest, the performance of male and female students in LB courses was similar to the EBAE courses. However, on the posttest, both male (female) students in the EBAE courses outperformed male (female) students in the LB courses (effect sizes ranging from 0.299 to 0.623). Interestingly, the effect sizes for male students were slightly higher than the effect sizes for female students, suggesting that male students may have benefited more from evidence-based pedagogies. The gender gap data shown in Table II can be interpreted in a similar manner. Thus, our data suggest that in calculus-based physics II, while both female and male students learned more in EBAE courses, male students may have benefited more than female students, resulting in a slight increase in the gender gap from pretest to posttest in EBAE courses.

Table IV. Between-course comparison of the average CSEM pre/posttest scores of calculus-based male and female students in LB courses with EBAE courses when (i) both courses are taught by the same instructor and (ii) different instructors using similar instructional methods are combined. The p-values and effect sizes are obtained for male and female students separately when comparing the LB and EBAE courses in terms of students' CSEM scores.

| | CSEM | Female-Pretest | Female-Posttest | Male-Pretest | Male-Posttest |
|---|---|---|---|---|---|
| (i) CSEM Calc: LB vs. EBAE (same instructor) | LB | N: 51 Mean: 35% SD: 12% | N: 44 Mean: 44% SD: 15% | N: 126 Mean: 42% SD: 12% | N: 110 Mean: 50% SD: 16% |
| | LB vs. EBAE Comparison | p-value: 0.590 Eff. size: 0.097 | p-value: 0.119 Eff. size: 0.299 | p-value: 0.845 Eff. size: 0.024 | p-value: 0.001 Eff. size: 0.455 |
| | EBAE | N: 75 Mean: 36% SD: 14% | N: 68 Mean: 48% SD: 17% | N: 133 Mean: 42% SD: 15% | N: 113 Mean: 58% SD: 20% |
| (ii) CSEM Calc LB vs. EBAE (different instructors combined) | LB | N: 84 Mean: 34% SD: 13% | N: 78 Mean: 45% SD: 16% | N: 234 Mean: 38% SD: 13% | N: 248 Mean: 51% SD: 17% |
| | LB vs. EBAE Comparison | p-value: 0.595 Eff. size: 0.077 | p-value: 0.003 Eff. size: 0.448 | p-value: 0.679 Eff. size: 0.039 | p-value: <0.001 Eff. size: 0.623 |
| | EBAE | N: 112 Mean: 35% SD: 14% | N: 98 Mean: 53% SD: 18% | N: 220 Mean: 39% SD: 16% | N: 193 Mean: 63% SD: 19% |

**RQ3. Comparison between EBAE and LB courses taught by the same instructor in terms of the performance of male and female students (divided according to the pretest scores)**

Tables V(a) and V(b) show the average algebra-based FCI and calculus-based CSEM pretest, posttest, gain, normalized gain and final exam scores for male and female students, along with the p-values between each subgroup of male and female students for pretest, posttest and the final exam in the EBAE and LB courses taught by the same instructor (with the same homework and final exam) with students divided into three subgroups based on their pretest scores. The male students were divided into three subgroups according to the pretest scores of male students only and female students were also divided into three subgroups according to the pretest scores of female students only. (Tables A1 and A2 in the appendix show similar type of information as Tables V(a) and V(b) except that the total number of students was divided

into three subgroups according to their pretest scores regardless of their gender and then male and female students were separated for comparison.)

**Algebra-based Physics I:** The data in Table V(a) show that in both the LB and EBAE algebra-based courses, on the pretest, there was a gender gap on the FCI between male and female students in each group (bottom 1/3, middle 1/3, top 1/3) and since female and male students had comparable gains, the gender gap was maintained in most cases on the posttest. This is consistent with the data shown in Table I which indicate that in both LB and EBAE courses, when including all students, the gender gap on the FCI stayed roughly the same from pretest to posttest. When comparing the LB with the EBAE course, we see that both female and male students seemed to benefit equally (gains and normalized gains on the FCI were higher in the EBAE course compared to the LB course), with the exception of the top 1/3 of the students. The top 1/3 of the female students had similar FCI gains in the LB and EBAE course (13% and 14%, respectively), whereas the top 1/3 of the male students showed larger FCI gains in the EBAE course.

Table V(a). Average FCI pretest scores (Pretest), posttest scores (Posttest), gain (Gain), normalized gain (Norm g) and final exam scores (Final) for male and female students in the EBAE and LB courses taught by the same instructor (with same homework and final exam). Male students were divided into three groups based upon their pretest scores and female students were also divided into three groups based upon their pretest scores separately. For each division (subgroup), a *p*-value was obtained using a *t*-test that shows whether there is statistically significant difference between male and female students on pretest, posttest and final exam.

| | Pretest Split | | Pretest | Posttest | Gain | Norm g | Final |
|---|---|---|---|---|---|---|---|
| | | Mean Female Score | 17 | 35 | 18 | 21 | 47 |
| | bottom 1/3 | p-value | 0.010 | 0.078 | | | 0.815 |
| | | Mean Male Score | 23 | 42 | 19 | 25 | 48 |
| FCI Alg LB | | Mean Female Score | 28 | 42 | 14 | 19 | 53 |
| | middle 1/3 | p-value | <0.001 | 0.012 | | | 0.078 |
| | | Mean Male Score | 39 | 55 | 16 | 26 | 60 |
| | | Mean Female Score | 44 | 57 | 13 | 23 | 60 |
| | top 1/3 | p-value | <0.001 | 0.002 | | | 0.422 |
| | | Mean Male Score | 65 | 74 | 8 | 24 | 64 |
| | | Mean Female Score | 15 | 39 | 23 | 27 | 54 |
| | bottom 1/3 | p-value | <0.001 | 0.164 | | | 0.318 |
| | | Mean Male Score | 21 | 43 | 22 | 28 | 50 |
| FCI Alg EBAE | | Mean Female Score | 28 | 46 | 17 | 24 | 52 |
| | middle 1/3 | p-value | <0.001 | <0.001 | | | 0.008 |
| | | Mean Male Score | 38 | 58 | 21 | 33 | 61 |
| | | Mean Female Score | 49 | 63 | 14 | 28 | 60 |
| | top 1/3 | p-value | <0.001 | <0.001 | | | 0.011 |
| | | Mean Male Score | 63 | 78 | 15 | 41 | 68 |

On the final exam, the data in Table V(a) suggest that in the EBAE class, the top and middle 1/3 of the male students performed better than the top and middle 1/3 of the female students. Similar, although not as strong, trends can be seen in the LB course. Since students in these two courses took the same final exams, these data suggest that the top and middle 1/3 of the male students benefited slightly more from EBAE pedagogies than top and middle 1/3 of the female students. On the other hand, the bottom 1/3 of the female students benefited more from EBAE pedagogies than the bottom 1/3 of the male students (performance of bottom 1/3 of female students is 54% in the EBAE course and only 47% in the LB course, whereas the performance of the bottom 1/3 of male students is 50% in the EBAE course and 48% in the LB course).

**Calculus-based Physics II:** The data in Table V(b) indicate that in the LB course, there was a gender gap on the CSEM in the pretest, but on the posttest, this gender gap decreased and was no longer statistically significant. This result appears to be inconsistent with the data shown in Table II which indicate that in the LB courses, the CSEM gender gap remained roughly the same (or slightly increased). However, the data in Table II includes all LB courses, whereas the data in Table V(b) includes only one course which was taught by the same instructor who taught the EBAE course shown in Table V(b). So in this particular calculus-based LB course, the gender gap on the CSEM decreased slightly, but if we include all calculus-based LB courses, the gender gap was roughly the same (or increased slightly).

Table V(b). Average CSEM pretest scores (Pretest), posttest scores (Posttest), gain (Gain), normalized gain (Norm g) and final exam scores (Final) for male and female students in the EBAE and LB courses taught by the same instructor (with same homework and final exam). Male students were divided into three groups based upon their pretest scores and female students were also divided into three groups based upon their pretest scores separately. For each division (subgroup), a *p*-value was obtained using a *t*-test that shows whether there is statistically significant difference between male and female students on pretest, posttest and final exam.

| | Pretest Split | | Pretest | Posttest | Gain | Norm g | Final |
|---|---|---|---|---|---|---|---|
| **CSEM Calc LB** | bottom 1/3 | Mean Female Score | 24 | 36 | 12 | 16 | 42 |
| | | p-value | 0.006 | 0.738 | | | 0.209 |
| | | Mean Male Score | 28 | 38 | 10 | 13 | 47 |
| | middle 1/3 | Mean Female Score | 34 | 44 | 10 | 16 | 52 |
| | | p-value | <0.001 | 0.661 | | | 0.770 |
| | | Mean Male Score | 40 | 46 | 6 | 10 | 50 |
| | top 1/3 | Mean Female Score | 49 | 56 | 8 | 15 | 53 |
| | | p-value | 0.060 | 0.513 | | | 0.018 |
| | | Mean Male Score | 54 | 59 | 5 | 12 | 62 |
| **CSEM Calc EBAE** | bottom 1/3 | Mean Female Score | 22 | 36 | 14 | 18 | 48 |
| | | p-value | 0.002 | 0.050 | | | 0.196 |
| | | Mean Male Score | 27 | 44 | 18 | 24 | 53 |
| | middle 1/3 | Mean Female Score | 34 | 48 | 14 | 22 | 52 |
| | | p-value | <0.001 | 0.455 | | | 0.104 |
| | | Mean Male Score | 41 | 51 | 10 | 18 | 58 |
| | top 1/3 | Mean Female Score | 56 | 63 | 7 | 16 | 68 |
| | | p-value | 0.382 | 0.003 | | | 0.621 |
| | | Mean Male Score | 58 | 74 | 16 | 38 | 70 |

For the calculus-based EBAE course, the data in Table V(b) suggest that the gender gap increased. The gender gaps for bottom 1/3, middle 1/3 and top 1/3 of the students are 5%, 7% and 2% in the pretest but 8%, 3% and 11% in the posttest, respectively. Thus, with the exception of the middle 1/3 of the students, the gender gap on the CSEM increased. This suggests that the bottom 1/3 and top 1/3 of the male students may have benefited more from EBAE pedagogies compared to the respective female students. It appears that this was indeed the case when we compare the normalized gains in the LB and EBAE courses: for the bottom 1/3 and top 1/3 of the male students, their CSEM normalized gains were 13% and 12% in the LB course, but 24% and 38% in the EBAE course, respectively. For the bottom 1/3 and top 1/3 of the female students, their CSEM normalized gains were 16% and 15% in the LB course, and 18% and 16% in the EBAE course. On the final exam, in both the LB and EBAE course, it appears that male students performed slightly better than the female students. However, only the 9% gender gap between the top 1/3 of the male and female students in the LB course is statistically significant.

**RQ4. Comparison between EBAE and LB courses taught by different instructors in terms of the performance of male and female students (divided according to the pretest scores)**

Tables VI(a) and VI(b) show the average algebra-based and calculus based FCI and CSEM pretest score, posttest score, gain and normalized gain for male and female students, along with the p-values between each subgroup of male and female students for the pretest and posttest in the EBAE and LB courses. All equivalent (algebra-based or calculus-based physics I or II) courses which used the same instructional strategy (EBAE or LB) were combined and students were divided into three groups based upon their pretest scores. Male students were divided into three subgroups according to the pretest scores of male students only and female students were also divided into three subgroups according to the pretest scores of female students only, and their scores were compared (cases in which male and female scores are significantly different have been highlighted). We note that Tables A3 and A4 in the appendix show the same data except that the students were divided into three subgroups according to their pretest scores

Table VI(a). Average FCI pretest scores (Pretest), posttest scores (Posttest), gain (Gain) and normalized gain (Norm g) for male and female students in the EBAE and LB algebra-based and calculus-based courses. All courses in the same group were combined. Male students were divided into three groups based upon their pretest scores and female students were also divided into three groups based upon their pretest scores separately. For each division (subgroup), a *p*-value was obtained using a *t*-test that shows whether there is statistically significant difference between male and female students on pretest and posttest. ***Note that FCI data for calculus-based EBAE classes are not available.***

| | Pretest Split | | Pretest | Posttest | Gain | Norm g |
|---|---|---|---|---|---|---|
| | | Mean Female Score | 24 | 39 | 15 | 19 |
| | bottom 1/3 | p-value | <0.001 | <0.001 | | |
| | | Mean Male Score | 35 | 52 | 17 | 26 |
| | | Mean Female Score | 43 | 52 | 9 | 16 |
| FCI Calc LB | middle 1/3 | p-value | <0.001 | <0.001 | | |
| | | Mean Male Score | 60 | 75 | 14 | 36 |
| | | Mean Female Score | 67 | 72 | 6 | 17 |
| | top 1/3 | p-value | <0.001 | 0.023 | | |
| | | Mean Male Score | 85 | 82 | -3 | -18 |
| | | Mean Female Score | 15 | 39 | 23 | 27 |
| | bottom 1/3 | p-value | <0.001 | 0.164 | | |
| | | Mean Male Score | 21 | 43 | 22 | 28 |
| | | Mean Female Score | 28 | 46 | 17 | 24 |
| FCI Alg EBAE | middle 1/3 | p-value | <0.001 | <0.001 | | |
| | | Mean Male Score | 38 | 58 | 21 | 33 |
| | | Mean Female Score | 49 | 63 | 14 | 28 |
| | top 1/3 | p-value | <0.001 | <0.001 | | |
| | | Mean Male Score | 63 | 78 | 15 | 41 |
| | | Mean Female Score | 17 | 33 | 15 | 18 |
| | bottom 1/3 | p-value | <0.001 | <0.001 | | |
| | | Mean Male Score | 23 | 41 | 18 | 23 |
| | | Mean Female Score | 29 | 44 | 15 | 22 |
| FCI Alg LB | middle 1/3 | p-value | <0.001 | <0.001 | | |
| | | Mean Male Score | 40 | 55 | 15 | 24 |
| | | Mean Female Score | 44 | 57 | 13 | 23 |
| | top 1/3 | p-value | <0.001 | <0.001 | | |
| | | Mean Male Score | 62 | 76 | 14 | 36 |

regardless of their gender. Then, male and female students were separated for comparison and the cases in which male and female scores are significantly different from each other have been highlighted. In these tables (VI(a), VI(b), A3 and A4), the average final exam performance is not listed because different instructors used different exams which varied in difficulty.

**Physics I:** For the calculus-based LB course, the data in Table VI(a) suggest that the gender gap on the FCI increased from pretest to posttest for each ability level. The gender gap for the bottom, middle, and top 1/3 of the students was 11%, 17%, 18% in the pretest, but in the posttest, it was 13%, 23%, 10%, respectively. This is consistent with the data in Table I which indicates that, including all students, the FCI gender gap increased slightly from the pretest to the posttest except the top 1/3 group.

Table VI(b). Average CSEM pretest scores (Pretest), posttest scores (Posttest), gain (Gain) and normalized gain (Norm g) for male and female students in the EBAE and LB algebra-based and calculus-based courses. All courses in the same group were combined. Male students were divided into three groups based upon their pretest scores and female students were also divided into three groups based upon their pretest scores separately. For each division (subgroup), a $p$-value was obtained using a $t$-test that shows whether there is statistically significant difference between male and female students on pretest and posttest. ***Note that CSEM data for algebra-based EBAE classes are not available.***

| | Pretest Split | | Pretest | Posttest | Gain | Norm g |
|---|---|---|---|---|---|---|
| | | Mean Female Score | 21 | 43 | 22 | 28 |
| | bottom 1/3 | p-value | 0.044 | <0.001 | | |
| | | Mean Male Score | 23 | 57 | 35 | 45 |
| CSEM Calc EBAE | | Mean Female Score | 32 | 52 | 20 | 29 |
| | middle 1/3 | p-value | <0.001 | 0.403 | | |
| | | Mean Male Score | 37 | 55 | 18 | 29 |
| | | Mean Female Score | 51 | 67 | 15 | 31 |
| | top 1/3 | p-value | 0.039 | 0.017 | | |
| | | Mean Male Score | 56 | 74 | 18 | 41 |
| | | Mean Female Score | 21 | 37 | 16 | 20 |
| | bottom 1/3 | p-value | 0.001 | 0.429 | | |
| | | Mean Male Score | 25 | 40 | 15 | 20 |
| CSEM Calc LB | | Mean Female Score | 32 | 43 | 12 | 17 |
| | middle 1/3 | p-value | <0.001 | 0.030 | | |
| | | Mean Male Score | 37 | 49 | 12 | 20 |
| | | Mean Female Score | 47 | 60 | 13 | 24 |
| | top 1/3 | p-value | 0.025 | 0.744 | | |
| | | Mean Male Score | 52 | 57 | 6 | 12 |
| | | Mean Female Score | 15 | 34 | 20 | 23 |
| | bottom 1/3 | p-value | 0.017 | 0.100 | | |
| | | Mean Male Score | 16 | 39 | 23 | 27 |
| CSEM Alg LB | | Mean Female Score | 22 | 40 | 19 | 24 |
| | middle 1/3 | p-value | <0.001 | 0.009 | | |
| | | Mean Male Score | 25 | 47 | 22 | 30 |
| | | Mean Female Score | 31 | 47 | 16 | 23 |
| | top 1/3 | p-value | <0.001 | <0.001 | | |
| | | Mean Male Score | 40 | 59 | 19 | 31 |

For the algebra-based LB and EBAE courses, the data in Table VI(a) suggest that the gender gap on the FCI was present at each ability level in the pretest and it remained roughly the same in the posttest (consistent with the data in Table I). For the LB courses, the gender gap on the FCI for bottom, middle, top 1/3 of the students was 6%, 11%, 18% on the pretest and 7%, 11%, 19% on the posttest. For the EBAE courses, the gender gap for bottom, middle, top 1/3 of the students was 6%, 10%, 14% on the pretest and 4%, 12%, 15% on the posttest. Interestingly, in both type of courses, it appears that the gender gap on the FCI was more pronounced at higher ability levels (based on FCI pretest scores). This was especially true in the LB course where the performance of the top 1/3 of the male students was on the average 18% (19%) higher compared to the top 1/3 of the female students on the pretest (posttest). The data in Table VI(a) suggest that both female and male students learned more in the EBAE course compared to the LB, but the learning gains were not much larger in the EBAE course compared to the LB course.

**Physics II:** The data in Table VI(b) suggest that in the calculus-based EBAE courses, the gender gap on the CSEM increased, but only for the bottom 1/3 of the students. On the pretest, the gender gap between the bottom 1/3 of the male and female students was 2%, whereas in the posttest, the gender gap was 14%. This suggests that the bottom 1/3 male students benefited much more from EBAE pedagogies than the bottom 1/3 of the female students. For the middle and top 1/3 of the students, the gender gap remained roughly the same. By comparison, in the calculus-based LB courses, the gender gap stayed roughly the same. The gender gap between bottom 1/3, middle 1/3, top 1/3 of the students was 4%, 5%, 5% in the pretest and 3%, 6%, -3% in the posttest. These findings are consistent with the data shown in Table II, which indicate that the gender gap on the CSEM stayed roughly the same in the LB courses, but increased slightly in the EBAE courses.

Comparing the LB with the EBAE courses in terms of normalized gain, the data in Table VI(b) suggest that students at all levels benefit from EBAE pedagogies. However, it appears that the bottom 1/3 and top 1/3 of the male students benefited more from EBAE pedagogies compared to the corresponding female students. The normalized CSEM gains for the bottom 1/3 and top 1/3 of the male students were 20% and 12% in the LB course but the EBAE course they were much larger at 45% and 41%. For the bottom 1/3 and top 1/3 of the female students on the other hand, the normalized CSEM gains were 20% and 24% in the LB course but only slightly larger at 28% and 31% in the EBAE course. A very similar trend was observed in Table V(b). Thus, it appears that for calculus-based physics II, the bottom 1/3 and top 1/3 of the male students may have benefited more from EBAE pedagogies compared to the bottom 1/3 and top 1/3 of the female students.

Similar to the calculus-based LB courses, for the algebra-based LB courses, the gender gap stayed roughly the same for students in the bottom, middle and top 1/3 of the class (based on CSEM pretest scores).

### RQ5. Correlation between CSEM posttest and final exam scores for male and female students

Figure 1 plots the CSEM posttest performance along with the final exam performance for male, female and all students in calculus-based EBAE course. Fig. 1 shows that the linear regressions [74] and there are moderate to strong correlation between posttest and the final exam scores. We also plotted linear regressions for the other courses and the data look similar to Fig. 1 but are not included here. Instead, we include all the correlation coefficients (CSEM posttest vs. final exam) for all the courses for which we were able to obtain both the posttest and final exam data. Table VII summarizes the correlation coefficients between CSEM/FCI posttest and final exam scores for each instructor who provided final exam data.

Despite the fact that different instructors had different final exams and at least some of the content in the final exam does not match the FCI or CSEM tests (e.g., in physics I, many topics were covered which are not on the FCI, such as momentum and collisions, static equilibrium and rotations, fluid dynamics and others), the correlation coefficients including males and females range from 0.415 to 0.816, which are considered to be moderate to high correlations. Although there are some differences between the correlation coefficients for male and female students in a given course, there is no clear discernable trend.

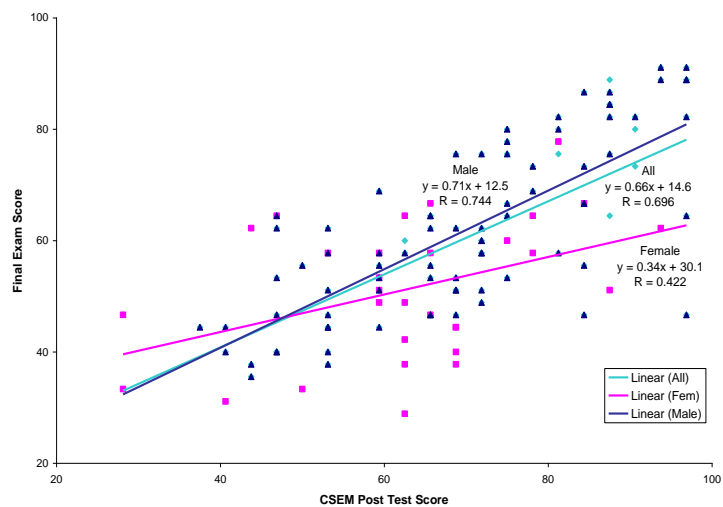

Fig 1. Linear regression and correlation coefficients of the CSEM posttest scores (conceptual) and final exam scores (heavy focus on quantitative problem solving) for male students, female students and all students (males and females combined together) for Inst 1 (EBAE instructor) for calculus-based introductory physics courses. The correlation coefficients for other FCI/CSEM instructors have been summarized in Table VII.

Table VII: Correlation coefficients (R) between CSEM/FCI posttest and final exam scores of male and female students for each instructor (Inst) who provided final exam data. The final exam data were not provided by physics II instructors in algebra-based courses.

| Physics I (Calc) | | | Physics I (Alg) | | | Physics II (Calc) | | |
|---|---|---|---|---|---|---|---|---|
| Inst and course | Female | Male | Inst and course | Female | Male | Inst and course | Female | Male |
| Inst 1: LB | 0.427 | 0.525 | Inst 1: EBAE | 0.568 | 0.628 | Inst 1: EBAE | 0.422 | 0.744 |
| Inst 2: LB | 0.674 | 0.573 | Inst 1: LB | 0.510 | 0.554 | Inst 2: EBAE | 0.571 | 0.415 |
| Inst 3: LB | 0.774 | 0.816 | Inst 2: LB | 0.626 | 0.728 | Inst 2: LB | 0.553 | 0.540 |
| | | | | | | Inst 3: LB | 0.230 | 0.592 |

## IV. DISCUSSION AND SUMMARY

**General findings for EBAE vs. LB courses without consideration of gender**: In all cases investigated, we find that on average, introductory students in the courses which made significant use of EBAE methods outperformed those in courses primarily taught using LB instruction on standardized conceptual surveys (FCI or CSEM) on the posttest even though there was no statistically significant difference on the pretest. This was true both in the algebra-based and calculus-based physics I (primarily mechanics) and 2 (primarily E&M) courses. Also, the differences between EBAE and LB courses were observed both among students who performed well on the FCI/CSEM pretest (given in the first week of classes) and also those who performed poorly, thus indicating that EBAE instructional strategies helped students at all levels. However, the typical effect sizes for the differences between equivalent EBAE and LB courses was between 0.23-0.49, which are small. Thus, the benefits of these EBAE approaches were not as large as one might expect to observe. There are many potential challenges to using EBAE instructional strategies effectively, including but not limited to:

- Content coverage. There is often a lot of content covered in introductory physics courses and it is challenging to cover the same amount of content while also including frequent active learning activities during class and students are expected to take responsibility for learning some those things outside of classes.

- Lack of student buy-in of EBAE pedagogies, which may result in lack of appropriate engagement with self-paced learning tools outside of class. It is therefore important for instructors to frame the course for students and discuss the various instructional approaches that will be used in a course and why they are expected to be beneficial for student learning. Providing data to the students that support the use of evidence-based active learning strategies [48] can be helpful, and when possible, including explicit discussions connecting students' and instructors' goals for taking the course can also be beneficial [76].

- Lack of student engagement with in-class active learning activities (e.g., clicker questions and group problem solving). Students may not recognize on their own that they will learn best if they engage with the in-class activities to the best of their ability. Therefore, ensuring that in-class activities help all students learn is important. Furthermore, since peer collaboration is exploited in many EBAE classes to enhance student learning, ensuring that these activities are designed and incentivized in a manner that not only fosters positive inter-dependence (success of one student is contingent on the success of the group) but also individual accountability (students are expected to show that they learned from working in a group) is essential [4-5].

- Large class sizes may be an impediment. One approach faculty used in flipped courses was to split the class in two (the instructor met with each group for only half of the time as compared to an LB course, his/her total contact hours with students remained the same), thus forming smaller class sizes. But even if the class size goes from 200 to 100 students by this process of breaking the class into two halves, it may still be challenging to manage the in-class activities effectively. Undergraduate or graduate teaching assistants need to be trained to effectively help in facilitating in-class activities. In group activities, students often work at different rates, so effective approaches need to be adopted to ensure that those who finish early can help others.

**Impact on gender gap:** We found that the EBAE courses did not result in reducing the gender gap. For algebra-based courses, students at all levels learned more in the EBAE courses; however, it appears that both female and male students benefited from evidence-based pedagogies equally and the gender gap present in the pretest was also found on the posttest. For calculus-based courses, our data suggest that male students actually benefited more from evidence-based pedagogies, which resulted in an increase in the gender gap from the pretest to the posttest. One hypothesis for why the gender gap was steady in algebra-based courses but grew in the calculus-based courses is that in the calculus-based courses there are significantly fewer women which can impact their sense of belonging and self-efficacy. These issues were not investigated in this study.

Previous research has also found that sometimes evidence-based pedagogies result in a reduction of the gender gap [54-55], while in other cases they do not [56]. The reasons for the gender gap even in the pretest are complex and some have attributed the persistence of gender gap to issues such as societal gender stereotypes, stereotype threat, high anxiety classes, lack of social belonging for women in physics classes, the culture promoting fixed intelligence mindset (with men having the innate ability to excel in subjects such as physics), and low self-efficacy [60-67].

Some have suggested that the gender gap found on conceptual assessments may at least in part be due to stereotype threat [58,61-64], and the extent to which the classroom environment is perceived as threatening by female students which in turn can depend on the instructor and the instructional design. For example, as discussed earlier, research suggests that, for high school female students, taking a physics test can create an 'implicit stereotype threat' and can degrade their performance [61]. Such threat may be present even when taking the FCI or CSEM test at the beginning of the semester as a pretest and lead to a gender gap in performance. Also, female students may have a lower sense of social belonging and low self-efficacy in a physics class due to societal stereotypes about who belongs in physics and who is capable of doing physics.

Some have suggested that classes which are not only collaborative but also emphasize collaboration and reduce competition (e.g., by not grading on a curve) are likely to be perceived more positively by female students and may partly be responsible for the reduced gender gap in Ref. [54]. Even EBAE classrooms can be characterized as high or low anxiety classes depending on the extent to which the instruction was designed to be inclusive and whether it explicitly focused on promoting a sense of belongingness, self-efficacy and growth mindset for all students. The extent to which the instructor plays an encouraging role to promote these positive motivational factors, and emphasizes that he/she is there as a guide to help all students succeed and also emphasizes that struggling is a stepping stone to success and should be viewed positively may also play a role in dispelling the negative impacts of societal gender stereotypes about physics that accumulate over a female student's lifetime. Since fixed mindset about innate intelligence can be a factor for the poor performance of female students, instructors should take advantage of research finding about the importance of promoting growth mindset [67] in their physics classes. In particular, research shows that students who believe that the brain is like a muscle, and intelligence is malleable and can increase with effort are more likely to persevere and perform better than those who think that intelligence is fixed [67]. Moreover, research suggests that mindset can be changed with a very short intervention [67]. Since according to the national data [77], fewer female students are likely to have taken challenging high school physics courses (e.g., Advanced Placement) before taking the college-level course, college EBAE courses which do not explicitly take into account these motivational factors may unknowingly create a high anxiety classroom environment for students who have less prior knowledge (who are more likely to be female students). For example, if students work in small groups in an EBAE course and some students in the group "show off" their knowledge and the instructional design does not promote a growth mindset, or the importance of hard work and persistence in learning physics, students who have taken less challenging physics course may have their self-efficacy issues exacerbated as opposed to reduced. Therefore, these motivational issues should be addressed in all physics classes as part of the instructional design to create inclusive classroom environment, as suggested in Ref. [78].

In summary, in order to enhance student learning in EBAE courses, it important not only to develop effective EBAE learning tools and pedagogies commensurate with students' prior knowledge but also to investigate how to implement them appropriately and how to motivate and incentivize their usage to get buy-in from students in order for them to engage with them as intended. Furthermore, reducing the gender gap on conceptual assessments is a challenging endeavor and evidence-based pedagogies may not be sufficient. In order to reduce gender gap, it may be useful to pay attention to other factors, e.g., improving the sense of belonging and self-efficacy of female students, improving their intelligence mindset (so that they do not think of male students as having an innate ability to excel in physics that they do not have and view intelligence as something that is malleable and can be cultivated by focus and effort), and reducing competition and emphasizing collaboration.

## V. ACKNOWLEDGEMENTS

We are grateful to all the faculty and students who helped with the study. We thank the US National Science Foundation for award DUE-1524575.

# VI. APPENDIX

Table A1. Average FCI pretest scores (Pretest), posttest scores (Posttest), gain (Gain), normalized gain (Norm g) and final exam scores (Final) for male and female students in the flipped and LB courses taught by the same instructor (with same homework and final exam) with students divided into three groups regardless of their gender based on their pretest scores. For each division (subgroup), a *p*-value was obtained using a *t*-test that shows whether there is statistically significant difference between male and female students on pretest, posttest or final exam.

| | Pretest Split | | Pretest | Posttest | Gain | Norm g | Final |
|---|---|---|---|---|---|---|---|
| | | Mean Female Score | 19 | 35 | 17 | 20 | 48 |
| | bottom 1/3 | p-value | 0.192 | 0.171 | | | 0.082 |
| | | Mean Male Score | 15 | 40 | 25 | 30 | 40 |
| | | Mean Female Score | 32 | 46 | 13 | 20 | 53 |
| FCI Alg LB | middle 1/3 | p-value | 0.967 | 0.955 | | | 0.772 |
| | | Mean Male Score | 32 | 45 | 13 | 19 | 54 |
| | | Mean Female Score | 49 | 62 | 13 | 26 | 64 |
| | top 1/3 | p-value | 0.046 | 0.121 | | | 0.934 |
| | | Mean Male Score | 56 | 70 | 13 | 31 | 64 |
| | | Mean Female Score | 16 | 40 | 24 | 28 | 54 |
| | bottom 1/3 | p-value | 0.149 | 0.610 | | | 0.438 |
| FCI Alg EBAE | | Mean Male Score | 18 | 42 | 24 | 29 | 51 |
| | | Mean Female Score | 31 | 45 | 14 | 21 | 52 |
| | middle 1/3 | p-value | 0.038 | 0.013 | | | 0.159 |
| | | Mean Male Score | 33 | 54 | 21 | 32 | 57 |

| | | | | | | |
|---|---|---|---|---|---|---|
| | | Mean Female Score | 53 | 71 | 18 | 38 | 63 |
| | top 1/3 | p-value | 0.091 | 0.209 | | | 0.212 |
| | | Mean Male Score | 58 | 75 | 17 | 40 | 67 |

Table A2. Average CSEM pretest scores (Pretest), posttest scores (Posttest), gain (Gain), normalized gain (Norm g) and final exam scores (Final) for male and female students in the flipped and LB courses taught by the same instructor (with same homework and final exam) with students divided into three groups regardless of their gender based on their pretest scores. For each division (subgroup), a *p*-value was obtained using a *t*-test that shows whether there is statistically significant difference between male and female students on pretest, posttest and final exam.

| | Pretest Split | | Pretest | Posttest | Gain | Norm g | Final |
|---|---|---|---|---|---|---|---|
| | | Mean Female Score | 27 | 37 | 11 | 15 | 44 |
| | bottom 1/3 | p-value | 0.945 | 0.505 | | | 0.921 |
| | | Mean Male Score | 26 | 35 | 8 | 11 | 44 |
| CSEM Calc LB | | Mean Female Score | 39 | 49 | 10 | 16 | 52 |
| | middle 1/3 | p-value | 0.701 | 0.298 | | | 0.688 |
| | | Mean Male Score | 39 | 45 | 7 | 11 | 53 |
| | | Mean Female Score | 55 | 62 | 7 | 15 | 57 |
| | top 1/3 | p-value | 0.507 | 0.613 | | | 0.578 |
| | | Mean Male Score | 53 | 59 | 6 | 13 | 59 |
| | | Mean Female Score | 25 | 41 | 16 | 21 | 50 |
| | bottom 1/3 | p-value | 0.542 | 0.480 | | | 0.436 |
| | | Mean Male Score | 24 | 44 | 20 | 26 | 53 |
| CSEM Calc EBAE | | Mean Female Score | 38 | 49 | 11 | 18 | 52 |
| | middle 1/3 | p-value | 0.433 | 0.909 | | | 0.288 |
| | | Mean Male Score | 39 | 49 | 11 | 17 | 57 |
| | | Mean Female Score | 57 | 63 | 6 | 14 | 69 |
| | top 1/3 | p-value | 0.949 | 0.015 | | | 0.869 |
| | | Mean Male Score | 57 | 73 | 16 | 36 | 69 |

Table A3. Average FCI pretest scores (Pretest), posttest scores (Posttest), gain (Gain) and normalized gain (Norm g) for male and female students in the flipped and LB algebra-based and calculus-based courses. All courses in the same group were combined with students divided into three groups regardless of their gender based upon their pretest scores. For each division (subgroup), a *p*-value was obtained using a *t*-test that shows whether there is statistically significant difference between male and female students on pretest and posttest. ***Note that FCI data for Calculus-based EBAE classes are not available.***

| | Pretest Split | | Pretest | Posttest | Gain | Norm g |
|---|---|---|---|---|---|---|
| | | Mean Female Score | 30 | 43 | 13 | 19 |
| | bottom 1/3 | p-value | 0.083 | 0.016 | | |
| | | Mean Male Score | 32 | 49 | 17 | 25 |
| FCI Calc LB | | Mean Female Score | 53 | 60 | 7 | 14 |
| | middle 1/3 | p-value | 0.049 | <0.001 | | |
| | | Mean Male Score | 55 | 71 | 16 | 36 |
| | | Mean Female Score | 83 | 84 | 1 | 7 |
| | top 1/3 | p-value | 0.918 | 0.707 | | |
| | | Mean Male Score | 82 | 82 | -1 | -3 |
| FCI Alg EBAE | bottom 1/3 | Mean Female Score | 16 | 40 | 24 | 28 |
| | | p-value | 0.149 | 0.610 | | |

| | | | Pretest | Posttest | Gain | Norm g |
|---|---|---|---|---|---|---|
| | | Mean Male Score | 18 | 42 | 24 | 29 |
| | middle 1/3 | Mean Female Score | 31 | 45 | 14 | 21 |
| | | p-value | 0.038 | 0.013 | | |
| | | Mean Male Score | 33 | 54 | 21 | 32 |
| | top 1/3 | Mean Female Score | 53 | 71 | 18 | 38 |
| | | p-value | 0.091 | 0.209 | | |
| | | Mean Male Score | 58 | 75 | 17 | 40 |
| FCI Alg LB | bottom 1/3 | Mean Female Score | 19 | 34 | 15 | 19 |
| | | p-value | 0.467 | 0.427 | | |
| | | Mean Male Score | 20 | 36 | 16 | 20 |
| | middle 1/3 | Mean Female Score | 32 | 46 | 13 | 20 |
| | | p-value | 0.617 | 0.501 | | |
| | | Mean Male Score | 33 | 47 | 15 | 22 |
| | top 1/3 | Mean Female Score | 50 | 64 | 14 | 27 |
| | | p-value | 0.002 | 0.005 | | |
| | | Mean Male Score | 56 | 71 | 15 | 34 |

Table A4. Average CSEM pretest scores (Pretest), posttest scores (Posttest), gain (Gain) and normalized gain (Norm g) for male and female students in the flipped and LB algebra-based and calculus-based courses. All courses in the same group were combined with students divided into three groups regardless of their gender based upon their pretest scores. For each division (subgroup), a *p*-value was obtained using a *t*-test that shows whether there is statistically significant difference between male and female students on pretest and posttest. ***Note that CSEM data for algebra-based EBAE classes are not available.***

| | Pretest Split | | Pretest | Posttest | Gain | Norm g |
|---|---|---|---|---|---|---|
| | | Mean Female Score | 23 | 44 | 22 | 28 |
| | bottom 1/3 | p-value | 0.571 | 0.001 | | |
| | | Mean Male Score | 22 | 57 | 35 | 45 |
| CSEM Calc EBAE | | Mean Female Score | 35 | 56 | 22 | 33 |
| | middle 1/3 | p-value | 0.771 | 0.891 | | |
| | | Mean Male Score | 35 | 56 | 21 | 32 |
| | top 1/3 | Mean Female Score | 54 | 65 | 10 | 23 |
| | | p-value | 0.718 | 0.033 | | |
| | | Mean Male Score | 55 | 72 | 17 | 38 |
| CSEM Calc LB | bottom 1/3 | Mean Female Score | 23 | 39 | 16 | 20 |
| | | p-value | 0.367 | 0.827 | | |
| | | Mean Male Score | 24 | 39 | 15 | 20 |
| | middle 1/3 | Mean Female Score | 35 | 46 | 11 | 17 |
| | | p-value | 0.833 | 0.535 | | |
| | | Mean Male Score | 35 | 48 | 13 | 20 |
| | top 1/3 | Mean Female Score | 51 | 61 | 10 | 21 |
| | | p-value | 0.966 | 0.460 | | |
| | | Mean Male Score | 51 | 58 | 7 | 14 |
| CSEM Alg LB | bottom 1/3 | Mean Female Score | 15 | 36 | 21 | 25 |
| | | p-value | 0.402 | 0.576 | | |
| | | Mean Male Score | 15 | 38 | 23 | 27 |
| | middle 1/3 | Mean Female Score | 23 | 42 | 20 | 25 |
| | | p-value | 0.981 | 0.192 | | |
| | | Mean Male Score | 23 | 46 | 23 | 30 |
| | top 1/3 | Mean Female Score | 33 | 45 | 13 | 19 |
| | | p-value | 0.002 | <0.001 | | |
| | | Mean Male Score | 37 | 56 | 19 | 30 |